\documentclass[prd,aps,preprint,epsfig,floats]{revtex4}
\usepackage{graphics}
\usepackage{epsfig}

\usepackage{graphicx}

\begin{document}

\preprint{\vbox{ \hbox{UMD-PP-06-008} }}
\title{\Large\bf $S_3$ Symmetry and Tri-bimaximal Mixing}
\author{\bf R.N. Mohapatra, S. Nasri and Hai-Bo Yu }

\affiliation{ Department of Physics, University of Maryland,
College Park, MD 20742, USA}

\date{May, 2006}

\begin{abstract}
The near maximal value for the atmospheric neutrino mixing angle
together with the fact that the solar mixing angle satisfies the
relation $\sin^2\theta_\odot\simeq \frac{1}{3}$ is the basis for the
so called tri-bimaximal mixing when $\theta_{13}=0$. In this
note, we explore the possibility that tri-bimaximal mixing is an
indication
of a softly broken higher leptonic symmetry $S_3$, the permutation
of three lepton families that embeds the $\mu-\tau$ exchange
symmetry of leptons.
 \end{abstract}
\maketitle
\section{Introduction}
Observation of nonzero neutrino masses and determination of two of
their three mixing parameters by experiments have raised the hope
that neutrinos may hold the key to unraveling the flavor puzzle
for quarks and leptons \cite{rev}. In order to make progress
towards realizing this goal, one must first decipher the
underlying reason for the observed leptonic mixing pattern and
then search for a unified description of quarks and leptons to
understand the quark flavor puzzle.

An interesting possibility to explore is that lepton mixings are
indications of underlying flavor symmetries. Two tantalizing hints
in favor of this are the near maximal atmospheric mixing angle and
vanishing mixing angle $\theta_{13}$. It has been speculated that
they are consequences of an approximate discrete $\mu-\tau$
symmetry\cite{mutau,moh} for leptons. Its presence can
be tested under certain circumstances. Even though there is no
such apparent ``$\mu-\tau$'' symmetry among quarks and charged
leptons, it has been shown that unified description of quarks and
leptons with this symmetry is possible\cite{nishiura}. A question
raised by this is whether there are higher underlying symmetries
of leptons.

A hint for a higher symmetry may be coming from the observation
that the solar angle in the PMNS mixing matrix satisfies the
relation $\sin^2\theta_\odot\simeq \frac{1}{3}$. The resulting
PMNS matrix has the simple form\cite{tbm}:
\begin{eqnarray}
U_{PMNS}~=~\pmatrix{\sqrt{\frac{2}{3}} & \frac{1}{\sqrt{3}} & 0\cr
-\frac{1}{\sqrt{6}} & \frac{1}{\sqrt{3}} & \frac{1}{\sqrt{2}}\cr
-\frac{1}{\sqrt{6}}& \frac{1}{\sqrt{3}} & -\frac{1}{\sqrt{2}}}
\end{eqnarray}
and is called a tri-bimaximal mixing. The true nature of the
symmetry responsible for this pattern is not clear, although there
are many interesting suggestions\cite{a4,o3,s3}.

In this brief note, we explore the possibility that the relevant
symmetry may be the permutation symmetry $S_3$ of three lepton
generations.
Such connections have been considered in literature from
phenomenological points of view\cite{s3} and it will be interesting
to see to what extent a full gauge model for tri-bimaximal pattern
based on $S_3$ group can be developed. In this note we take some steps
in this direction. We show that a softly broken $S_3$ symmetry for
leptons can lead to tri-bimaximal mixing pattern if we use a
combination of type I and type II seesaw mechanism to understand the
smallness of neutrino mass. This approach appears to be different
from previous attempts at building models for tri-bimaximal
mixing\cite{a4,o3,s3}.

 We proceed in two steps: we first
show how in a basis where charged leptons are diagonal, one can
derive the mixing pattern in Eq. (1) using softly broken $S_3$
symmetry under certain assumptions. We then show how this the $S_3$
symmetry combined with $ Z_{2e}\times Z_{2\mu}\times Z_{2\tau}$
symmetry can lead to a diagonal charged lepton mass matrix. We then
extrapolate the neutrino mass matrix from the seesaw scale to the
weak scale and obtain constraints on the mass ratios $m_1/m_3$ and
$m_2/m_3$ so that the mixing angles match the observations. We
obtain a prediction for $\theta_{13}$, which turns out to be
extremely small ($\sim 0.004$). We further show that if the neutrino
masses are quasi-degenerate and have the same CP property (i.e. are all
positive), then the radiative corrections in the extrapolation to
the weak scale are so large that the solar mixing angle is in disagreement 
with observations. This
implies that in supersymmetric theories with large tan$\beta$,
seesaw scale tri-bimaximal mixing and degenerate neutrinos are
mutually exclusive.

\section{An $S_3$ model}

We start with the Majorana neutrino mass matrix whose
diagonalization at the seesaw scale leads to the tri-bimaximal
mixing matrix:
\begin{eqnarray}
{\cal M}_\nu~=~\pmatrix{a & b & b\cr b & a-c & b+c\cr b & b+c &
a-c}
\end{eqnarray}
Diagonalizing this matrix leads to the $U_{PMNS}$ of Eq. (1) and
the neutrino masses: $m_1=a-b; m_2=a+2b$ and $m_3=a-b-2c$. Clearly
if $|a|\simeq |b| \ll |c|$, we get a normal hierarchy for masses.

We now show that the mass matrix in Eq.(2) can be obtained from a
softly broken $S_3$ symmetry in the neutrino sector. For this
purpose, we assign the three lepton doublets of the standard model
$(L_e,L_\mu,L_\tau)$ to transform into each other under permutation.
The three  right handed neutrinos $(\nu_{R,i=1,2,3})$ transform
under three permutation and two cyclic operations of $S_3$ as:
\begin{eqnarray}
\nonumber e\leftrightarrow \mu:~ \nu_{R,1}\leftrightarrow -\nu_{R,1}
;~ \nu_{R,2}\leftrightarrow -\nu_{R,3}\\ \nonumber\mu\leftrightarrow
\tau:~ \nu_{R,2}\leftrightarrow -\nu_{R,2};~\nu_{R,1}\leftrightarrow
-\nu_{R,3} \\\nonumber \tau\leftrightarrow e:~
\nu_{R,3}\leftrightarrow -\nu_{R,3};~\nu_{R,1}\leftrightarrow
-\nu_{R,2}\\
\nonumber e\rightarrow\mu\rightarrow\tau:~
\nu_{R,1}\rightarrow\nu_{R,2};~\nu_{R,2}\rightarrow\nu_{R,3};~
\nu_{R,3}\rightarrow\nu_{R,1}\\
e\rightarrow\tau\rightarrow\mu:~\nu_{R,1}\rightarrow\nu_{R,3};~
\nu_{R,2}\rightarrow\nu_{R,1};~\nu_{R,3}\rightarrow\nu_{R,2}
\end{eqnarray}
In order to obtain the neutrino mass matrix, we assume that there
is a standard model triplet Higgs field $\Delta$ with $Y=2$ which
is $S_3$ singlet that couples to the two lepton doublets and  an
$S_3$ singlet Higgs doublet field $H$ that gives the Dirac mass
for the neutrinos. The triplet vev can be made small and of the desired
order if the mass of the triplet Higgs field is around $10^{14}$ GeV or
so\cite{rev}.

The first point to note is that the most general $S_3$ invariant
coupling of the triplet i.e. $f_{ab}L_aL_b \Delta$ is given by the
coupling matrix:
\begin{eqnarray}
f~=~\pmatrix{a & b & b\cr b & a & b\cr b & b & a}
\end{eqnarray}
For the Dirac neutrino coupling we choose to keep the following
$S_3$ invariant term:
\begin{eqnarray}
{\cal
L}_D~=~h_\nu[\overline{\nu_{R,1}}H(L_e-L_\mu)+
\overline{\nu_{R,2}}H(L_\mu-L_\tau)+\overline{\nu_{R,3}}H(L_\tau-L_e)]+h.c.
\end{eqnarray}
One other $S_3$ invariant coupling is set to zero. This
is natural in a supersymmetric theory due to the
nonrenormalization theorem.  We then get for the Dirac mass matrix
for neutrinos
\begin{eqnarray}
M_D=\left[\matrix{d&-d&0\cr 0&d&-d\cr -d&0&d}\right].
\end{eqnarray}
where $d=h_\nu<H>$. If we now assume the following hierarchy among
the right handed neutrinos, i.e. $M_{\nu_{R, 1,3}}\gg M_{\nu_{R,2}}$
so that a single right handed neutrino dominates the type I
contribution to the seesaw formula\cite{king}, then in the strict
decoupling limit, using the mixed type I+II seesaw formula:
\begin{eqnarray}
{\cal M}_\nu~=~M_0-M^T_DM^{-1}_{\nu_R}M_D,
\end{eqnarray}
we get the desired form for the neutrino Majorana mass matrix
(Eq.(2)) which leads to tri-bimaximal mixing. Note that the right handed
neutrino masses being dimension three operators break the $S_3$ softly.

In this discussion we have assumed that the charged lepton mass
matrix is diagonal. A major challenge for any model for neutrino
mixings is to have a consistent picture for both the charged lepton
and neutrino sectors simultaneously so that the combination
$U^{\dagger}_\ell U_\nu$ equals the observed PMNS matrix. Since in
our case, the neutrino sector by itself gives the tri-bimaximal form
for the PMNS matrix, the charged lepton sector should be diagonal or
nearly so. We will now show that we can obtain a diagonal charged
lepton mass matrix in a simple way using the $S_3$ symmetry,
provided we choose only one of two allowed $S_3$ invariant Yukawa
coupling terms.

In order to achieve this, we assume that there are three standard
model Higgs doublets $(H_e,H_\mu,H_\tau)$ transforming like the
lepton doublets above under $S_3$. We also assume that the right
handed charged leptons $(e_R,\mu_R, \tau_R)$ transform under $S_3$
same way. We then assume a product of discrete symmetries
$Z_{2e}\times Z_{2\mu}\times Z_{2\tau}$ under which all fields
except the following are even: $(e_R,H_e)$ odd under only $Z_{2e}$
and similarly $(\mu_R, H_\mu)$ are odd only under $Z_{2\mu}$ and
$(\tau_R, H_\tau)$ odd under $Z_{2\tau}$. The Yukawa couplings
invariant under this are:
\begin{eqnarray}
\nonumber {\cal L}'_Y~=~h_e (\bar{L}_eH_ee_R +\bar{L}_\mu
H_\mu\mu_R+\bar{L}_\tau
H_\tau\tau_R)~+~h'_e(\bar{L}_eH_\mu\mu_R+\bar{L}_\mu H_e
e_R\\+\bar{L}_\mu H_\tau\tau_R+\bar{L}_\tau
H_\mu\mu_R+\bar{L}_eH_\tau\tau_R+\bar{L}_\tau H_e e_R)+h.c.
\end{eqnarray}
By softly breaking the global $S_3$ symmetry in the Higgs potential
for the $H_{e,\mu,\tau}$, we can  get $<H_e>\ll <H_\mu>\ll <H_\tau>$
which allows us to obtain a realistic diagonal charged lepton mass
matrix if we assume $h'_e=0$. This model then gives us a
tri-bimaximal neutrino mixing at the seesaw scale.

\section{Some implications}
In order to compare this model with observations, we need
 to extrapolate the seesaw scale neutrino mass matrix in Eq.(2)
down to the weak scale\cite{rod} and then calculate the masses and
mixing angles. This extrapolation depends on the mass hierarchy of
the neutrinos. So comparing with observations, we can put limits on
the mass hierarchy at low scale. From the expressions for the
neutrino masses derived after Eq.(2), one might think that
degenerate masses are compatible with tri-bimaximal pattern since
there are three parameters and three masses to be fitted. However,
in supersymmetric models, mixing angles can receive substantial
contributions from RGE effects (specially for large tan$\beta$) and
will in general lead to distortion of the mixing angles away from
the tri-bimaximal values.  For the specific case of tan$\beta=50$
 we calculate the radiative
corrections to the solar mixing angle $\theta_{12}$ in Fig. 1. 
We plot sin$^2\theta_{12}$ against $m_2/m_3$ with the input constraint 
being that
$\Delta m^2_\odot/\Delta m^2_{ATM}$ is within 3 $\sigma$ of its present 
value i.e. $0.024\leq\Delta m^2_{\odot}/\Delta m^2_{ATM}\leq 
0.060$\cite{valle}. 
We see that for $m_2/m_3 > 0.3$ or so, the solar mixing angle 
goes outside the observed range and the agreement gets worse for larger 
values of this mass ratio which corresponds to quasi-degenerate neutrino 
spectrum. This leads us to conclude that tri-bimaximal mixing at the 
seesaw scale is incompatible with quasi-degenerate neutrinos for large 
values of $ \tan \beta$.

For the same value of $\tan \beta$, we show in Fig.2  the allowed
ranges for the neutrino mass ratios for the case of normal
hierarchy and in Fig.3, the prediction for $\theta_{13}$ for
this model. In these figures, we have used the above $3\sigma$ 
experimental 
bounds for $\Delta m^2_{\odot}/\Delta m^2_{ATM}$ and also 3 $\sigma$ 
bounds 
for $0.23\leq\sin\theta^2_{12}\leq 0.38$ and $0.34\leq\sin\theta^2_{23}\leq
0.68$\cite{valle}. We find that
the prediction for $\theta_{13}\sim 0.004$ which is much too small
to be observable in near future. This is because the low energy
theory in the absence of radiative corrections is $\mu-\tau$
symmetric.   Clearly observation of $\theta_{13}$ higher than this
value will rule out this model and indeed any simple model
for tri-bimaximal mixing at the seesaw scale for the case of normal mass
hierarchy.

\begin{figure}[h!]
\includegraphics[scale=0.9]{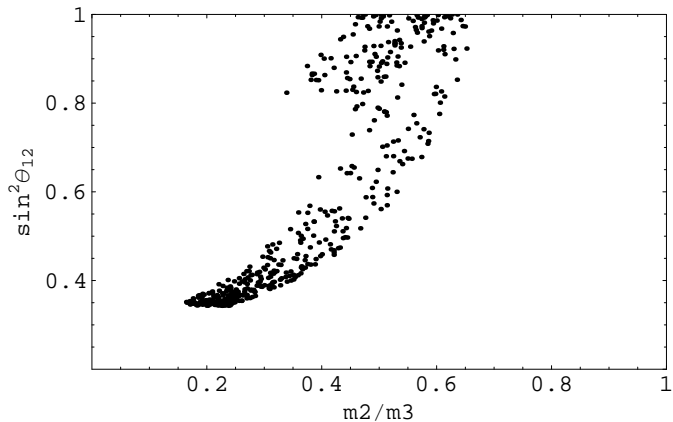}
\caption{sin$^2\theta_{12}$ at the weak scale for the case of quasi- 
degenerate neutrinos. Note that the higher the ratio $m_2/m_3$, the more 
degenerate the neutrinos are and further off the prediction for 
sin$^2\theta_{12}$ is from the observed value.} \end{figure}

\begin{figure}[h!]
\includegraphics[scale=0.9]{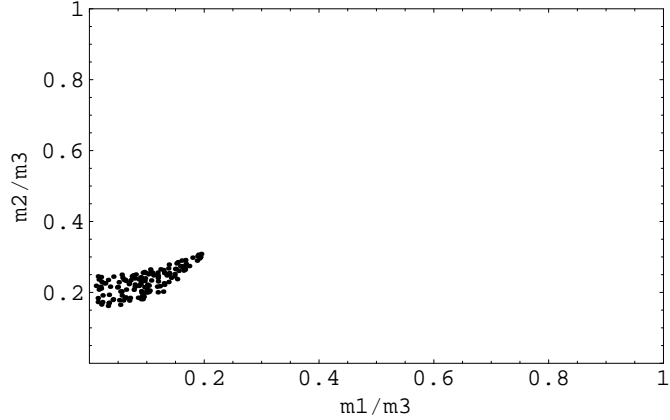}
\caption{Allowed ranges of mass ratios at the weak scale for normal
hierarchy case.}
\end{figure}

\begin{figure}[h!]
\includegraphics[scale=0.9]{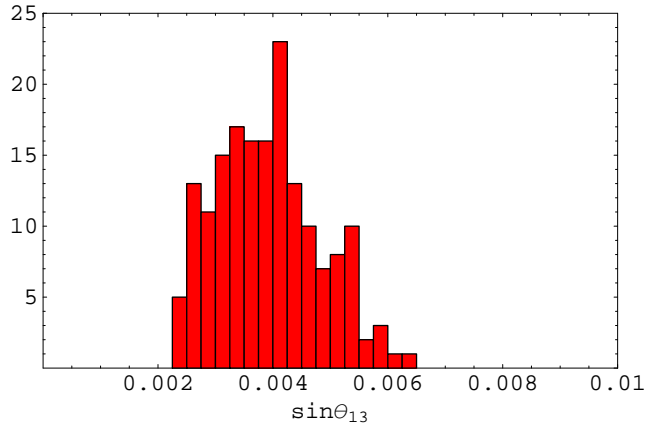}
\caption{Distribution of $\sin\theta_{13}$ value.}
\end{figure}

\section{Conclusion}
In conclusion, in this brief note, we have presented a new way to
obtain the tri-bimaximal mixing pattern for neutrinos by embedding
$\mu-\tau$ symmetry of the neutrino mass matrix into a softly broken
$S_3$ permutation symmetry for leptons and using a simple
combination of the type I and type II seesaw formulae along with the
dominance of a single right handed neutrino\cite{king}. We also find
that tri-bimaximal mixing at the seesaw scale is incompatible with
degenerate neutrino spectrum due to large radiative correction
effects for large $ \tan \beta$.

 This work is supported by the National Science Foundation grant
no. Phy-0354401

\end{document}